\documentclass[twocolumn, 12pt]{autartsofie}  
\usepackage{amsfonts,dblfloatfix,mathtools}
\usepackage{graphicx,subcaption}
\usepackage{graphics}

\usepackage{tikz}
\usepackage{color} 
\usepackage{pgfplots}
\usepackage{paralist}
\usepackage[nonumberlist]{glossaries}
\usepackage{amsmath,amssymb,amsfonts} 	 \usetikzlibrary{decorations.shapes,shapes.geometric,shadows,arrows,automata,positioning,calendar,mindmap,backgrounds,scopes,chains,er,patterns}
 
\usetikzlibrary{calc} 
\newtheorem{theorem}[thm]{Theorem}
\newtheorem{remark}{Remark}
\newglossary[slg]{symbolslist}{syi}{syg}{List of symbols}
 \newcommand{\mathscr}[1]{\mathcal{#1}}

 \newcommand{\colored}[1]{{ #1}}
 \newcommand{\colto}[1]{\textcolor{orange}{ #1}}

\newenvironment{proof}{\textbf{Proof}%
}{%
\hfill$\Box$%\end{flushright}
}
\allowdisplaybreaks

    \newenvironment{itemize*}%
  {\begin{itemize}%
    \setlength{\itemsep}{1pt}%
    \setlength{\parskip}{1pt}}%
  {\end{itemize}}
  \newenvironment{enumerate*}%
  {\begin{enumerate}%
    \setlength{\itemsep}{1pt}%
    \setlength{\parskip}{1pt}}%
  {\end{enumerate}}
 \newcommand{\mb}[1]{\mathbf{#1}}
 \newcommand{\mc}[1]{\mathcal{#1}} 
 \newcommand{\newsymb}[4]{
 \newcommand{#2}{#3}
 \newglossaryentry{symb:#1}{
 name=$#2$,
 description={#4},
 sort=symbol#1, type=symbolslist
 }
 }
 \newcommand{\sym}[1]{\gls{symb:#1}}
\newsymb{Hilbert}	{\Hilbert}	{\mathcal{H}_2}
{The Hilbert space} 
\newsymb{Reals}		{\Real}		{\mathbb R}
{The set of reals}        
\newsymb{Naturals}	{\Nat}		{\mathbf{N}}
{The set of natural numbers}             
\newsymb{Complex}	{\Complex}	{\mathbb C}  
{The set of complex numbers}
{} % normal distribution

\newcommand{\norm}[1]{\left\Vert#1\right\Vert}
\newsymb{norm}{\normempty}{\norm{\cdot}}  
{The Euclidean norm $\|x\|=\sqrt{x^Tx}.$}

\newcommand{\eps}{\varepsilon}  
 
\newsymb{system}{\s}{\mathbf{M}}{The true `(cyber-)physical' system}
\newsymb{model}{\M}{\mathbf{M}}{A mathematical model}
\newsymb{detmodel}{\Md}{\bar{\mathbf{M}}}{The minimal realisation of $\M$ when the noise input is neglected.}

\newsymb{controlmodel}{\Mc}{\mathbf{M}_\mathbf{C}}{The model composed with the controller $\mathbf{C}$}
\newsymb{controller}{\C}{\mathbf{C}}{The controller.}

\newsymb{statespace}{\X}{\mathbb{X}}{The state space, i.e. the set of states}     % State space
	\newsymb{state}{\xv}{\mathbf{x}}{An element of the state space $\X$}   
\newsymb{actionspace}{\A}{\mathbb{U}}{The set of controllable inputs}    
	\newsymb{actions}{\acv}{u}{An element of the set of controllable input $\A$}   
\newsymb{outputspace}{\Y}{\mathbb{Y}}{The set of output values}     % Observation space
    \newsymb{output}{\yv}{y}{An element of the set of outputs}
\newsymb{performanceoutput}{\Z}{\mathbb{Z}}{The set of performance outputs}     % Observation space

\newsymb{observer}{\Obs}{\mathbf{O}}{An observer}

\newsymb{TS}{\TS}{\Sigma}{A transition system}
	\newcommand{\I}[1]{{{#1}_a}} % example a
	\newcommand{\II}[1]{{{#1}_b}} % example b

	\newsymb{XTS}{\XTS}{\mathscr{X}}{The set of states of a transition system}
	\newsymb{xTS}{\xTS}{x}{The set of states of a transition system}
	\newsymb{UTS}{\UTS}{\mathscr{A}}{The set of inputs to a transition system} 
	\newsymb{uTS}{\uTS}{u}{The set of inputs to a transition system}
	\newsymb{ZTS}{\ZTS}{\mathscr{Z}}{The set of possible infinite observations}
	 \newsymb{HTS}{\HTS}{\mathscr{H}}{The labelling map}
	
\newsymb{initialstates}{\init}{\mathbb{I}}{The set of initial states}

\newsymb{Markov}{\Markov}{\mathcal{M}}{A Markov process}
\newsymb{Relation}{\rel}{\mathcal{R}}{A relation}
\newsymb{EquivRelation}{\erel}{\mathcal{R}}{An equivalence  relation}

\newsymb{similated}{\simu}{\preceq_{\mathcal{S}}}{Simulated by}
\newsymb{bisimilar}{\bsimu}{\sim_{\mathcal{B}}}{Bisimulation equivalence}
\newsymb{simulationequiv}{\esimu}{\simeq_{\mathcal{S}}}{Simulation equivalence}
\newsymb{altsimilated_app}{\asimu}{\preceq^\eps_{\mathcal{AS}}}{$\eps$-approximate alternating simulation}
\newsymb{similated_app}{\simuapp}{\preceq^\eps_{\mathcal{S}}}{$\eps$-approximate simulated by}

\newsymb{until}{\un}{\mbox{\textbf{U}}}{until operator}
\newcommand{\w}{w}
\definecolor{color1}{RGB}{0,0,90} % Color of the article title and sections
\definecolor{color2}{RGB}{0,20,20} % Color of the boxes behind the abstract and headings

\newcommand{\QEDopen}{\hfill\qed}

\begin{document} 
%\begin{centering}
%\hspace{2cm}
%\begin{minipage}[b]{1.4\linewidth}
%\textbf{\large Cover Letter}\\*[1.4em]
%We send this manuscript entitled ``Observer-based correct-by-design controller'' to be considered for publication in ... .\\
%\smallskip
%
%The design of controllers which verify properties encoded in a formal language is allows for safe and reliable controller synthesis. 
%In this work we give a synthesis method for output-feedback controllers which are correct by construction.
%\medskip
%
%This work is an extended version of the ideas first published in CDC. It includes the proofs omitted from publication to CDC.
%\medskip
% 
%
%\end{minipage}\end{centering}

\begin{frontmatter}
\title{Observer-based correct-by-design controller synthesis}
         \author[TUE]{S.Haesaert}, 
         \author[TUE]{P.M.J.Van den Hof}, 
         \author[OX]{A.Abate}
\address[TUE]{Department of Electrical Engineering, %\\
       Eindhoven University of Technology, %\\
       Eindhoven, The Netherlands}        
\address[OX]{Department of Computer Science, %\\
       University of Oxford, %\\
       Oxford, United Kingdom
       }

\begin{abstract}
Current state-of-the-art correct-by-design controllers are designed for full-state measurable systems. 
This work first extends the applicability of correct-by-design controllers to partially observable LTI systems.
Leveraging 2nd order bounds we give a design method that has a quantifiable robustness to  probabilistic disturbances on state transitions and on output measurements. 
%Using linear matrix inequalities we formulate this design problem as a convex optimisation problem.
In a case study from smart buildings we evaluate the new output-based correct-by-design controller on a physical system with limited sensor information. 
\\*
\end{abstract}
\end{frontmatter}

\setcounter{page}{1}
\section{Introduction}
\label{Introduction} 
             
Reliable and autonomous operation of many complex engineering systems demands guaranteed behaviour over the full spectrum of operating conditions. 
This is the case with applications in avionics, automotive, transportation systems, dependable electronics, semiconductors \cite{Vardi2006}, 
and in general in systems where safety is critical and where mistakes lead to impactful economical losses.   

Within the computer sciences, 
verification and synthesis of critical hardware and software has been attained in the industrial practice by tools and techniques from the domain of formal methods \cite{Clarke2008}.  
Employing well-structured specifications, 
such as properties expressed over linear-time temporal logics (LTL), 
automated and computer-aided tools have been developed for the verification and synthesis of models of the systems of interest. 
To meet new demands from domains dealing with complex new applications, 
these methods need to be extended to be applicable on models of (cyber-)physical systems. 
Recent research \cite{tabuada2006linear,Tabuada2009b,Jr2010}  
pursues this overall objective via the verification 
%and controller synthesis 
of models (of physical systems) with uncountable state spaces:    
of special interest is the safe-by-construction automatic synthesis of controllers. 
% \note{By abstracting the dynamics of the \colored{system}s a discrete state \colto{model} is found that represents the behaviour of the physical \colored{system} symbolically and for which the control problem can be solved in a certifiable manner. Using an equivalence relation between the symbolic \colto{model} and the \colored{system} the synthesised controller for the symbolic \colto{model} is refined to the original \colto{model} by lifting it  over the equivalence relation.}
These correct-by-design controllers are however incompatible with general systems for which models with exact knowledge of the dynamics and full state measurements are not available.    
\subsubsection*{Contributions}
In this work we extend correct-by-design controllers for linear time invariant (LTI) models as in \cite{tabuada2006linear} 
to output-based controllers that employ sensor outputs or partial state measurements. 
As in \cite{tabuada2006linear}, 
our new control architectures come with quantitative certificates on the accuracy. 
Further, 
since dynamics of physical systems are often disturbed in a probabilistic sense and associated sensors are noisy, 
we require the new output-based controllers to show quantifiable robustness with respect to stochastic disturbances on state evolutions and output measurements. 

\subsubsection*{Related work}
Design methods for classical optimal control problems 
%For the tuning and optimal control in classical control problems 
\cite{franklin1998digital} of models with (noisy) output measurements  
can be distinguished in direct designs based on the input-output behaviour of the system, 
and in methods exploiting the separation of estimation and control. 
The former class includes frequency-domain 
%such as PID tuning
and robust control methods;   
alternatively, 
whenever applicable (as in the optimal linear quadratic Gaussian problem) the separation theorem \cite{Witsenhausen1971} allows for the distinct design of an observer estimating the state and of a state feedback controller, 
yielding a combined output feedback controller.  
%Examples include the standard separation theorem for optimal linear quadratic Gaussian control design. 

Within the formal methods literature, 
limited efforts have targeted the synthesis of controllers for finite state models without state observations. 
Existing results target finite-state models:  
%This includes the in PRISM implemented 
\cite{Giro2012} studies the synthesis for partially observable models by searching the space of output-feedback controllers via counter-example-guided refinements. %\url{http://www.prism\colto{model}checker.org/bibitem.php?key=GR12}])} %  Visitor from oxford start from solution of synthesis problem for fully observable \colored{system}, refine to partial observability (Sergio \url{http://www.prism\colto{model}checker.org/bibitem.php?key=GR12})
A heuristic algorithm in \cite{Chatterjee2014} finds controllers satisfying LTL properties almost surely over partially-observable Markov decision processes. 
%Belta HSCC to appear 2015 (Chatterjee's work): M.Svorenova, M.Chmelik, K.Leahy, H.Ferit Eniser, K.Chatterjee, I.Cerna and C.Belta, ``Temporal Logic Motion Planning using POMDPs with Parity Objectives", Hybrid \colored{system}s: Computation and Control (HSCC) 2015 (to appear) 
In contrast, the work of \cite{zhang2005logic} extends PCTL$^\ast$ to target hidden Markov models and proposes a model checking algorithm. % Zhang, L. and Hermanns, H. and Jansen, D.N. (2005) Logic and {model} checking for hidden Markov {model}s. In: Formal techniques for networked and distributed \colored{system}s, FORTE (pp. pp. 98-112). \url{http://doc.utwente.nl/54796/}

For fully observable Markov Processes with general state spaces, 
verification and controller synthesis problems are reviewed in \cite{Abate2011}, 
and generally tackled over a simplified {model} that can be formally related to the original one. 
The simplified model can then shown to be in an (approximate) relation with the original model, 
either via metrics defined over the marginals of the conditional kernels \cite{sa2011}, 
or via metrics bounding the distance between the output trajectories \cite{Julius2009}. 
In contrast, 
this work will use the definition of approximate bisimulation relations, similar to those in \cite{Zamani}, to quantify the \emph{expected} deviation of noisy trajectories affected by stochastic disturbances.
 
%\subsubsection*{Contributions}
% In this work we extend the available work in two ways, first we extend correct-by-design controllers to the output feedback case. 
%Secondly we show that this approach is also viable for stochastic systems that can be modelled as linear time invariant systems with stochastic disturbances. \smallskip
 
\subsubsection*{Structure of the article}  
After reviewing preliminary notions in Sec. \ref{sec:2}, the problem statement together with state-based, correct-by-design controller architectures \cite{tabuada2006linear,Tabuada2009b} is given in Sec. \ref{prob.}.  
We design an output-based controller by introducing a state observer and a notion of output-based interface in Section \ref{sec:3}. 
Under very standard controllability and observability conditions on the model, 
this design allows us to bound the deviation between state-based and output-based controllers (cf. Sec \ref{sec:dev}).  
Additionally Section \ref{sec:4} discusses robustness issues with respect to stochastic disturbances, 
both on state transitions and sensor measurements. 
%, is quantified as the expected deviation of the output trajectories.
% the selection of the remaining free design parameters is discussed with respect to these disturbances.}
Finally in Section \ref{sec:5} the design methodology is evaluated on a case study in the area of Smart Buildings.

\section{Preliminaries} \label{sec:2}

\subsection{Transition systems and simulation relations} 
\begin{defn}[Transition {system} \cite{Tabuada2009b}]
\label{def:trans}
A transition {system} is a tuple \mbox{$\TS=(\XTS,\XTS_0,\UTS,\rightarrow,\ZTS,\HTS)$}, 
where
\begin{itemize*}
\item $\XTS$ is a (possibly infinite) set of states; % with $\xTS\in\XTS$; 
\item $\XTS_0$ is a (possibly infinite) set of initial states; 
\item $\UTS$ is a (possibly infinite) set of actions; 
\item $\rightarrow\subseteq \XTS\times \UTS\times  \XTS$ is a transition relation; 
\item $\ZTS$ is a (possibly infinite) set of observations; 
\item $\HTS:\XTS\rightarrow \ZTS$ is a map assigning to each $x\in \XTS$ an observation $\HTS(x)\in \ZTS$. \end{itemize*}
A metric transition {system} is a transition {system} endowed with a metric over the observation space $\ZTS$.\hfill\QEDopen
\end{defn} 
This work considers non-blocking transition {system}s, 
where every state $x\in \XTS$ is associated to a non empty transition relation. 
The behaviour generated by $\TS$ is denoted as $\mc B(\TS)$ and consists of all infinite sequences $z_0,z_1,z_2,...$ for which there exists an initialised path $(x_0,u_0),(x_1,u_1),(x_2,u_2),\ldots$, 
with $x_0\in \XTS_0$, $(x_i,u_i,x_{i+1})\in \rightarrow$, and $z_i=\HTS(x_i)$ for all $i\in \mathbb{N}$.

A transition system is called \emph{deterministic}  if the initial state is defined deterministically, i.e.,  $\XTS_0:=\{x_0\}$, and  for a given state-action pair the next state is determined uniquely. 

{The verification of LTI models can be attained by abstracting them as finite-state ones and  
leveraging symbolic approaches \cite{Tabuada2009b}. Pairs of models can be related as follows. } 

\begin{defn}[Simulation relation \cite{Tabuada2009b}]\label{def:sim} \mbox{ }\\
Let \mbox{$\I{\TS}=(\I{\XTS},\I{\XTS}_0,\I{\UTS},\I{\rightarrow},\I{\ZTS},\allowdisplaybreaks\I{\HTS})$} % \\* 
and  \\ $\II{\TS}=(\II{\XTS},\II{\XTS}_0,\II{\UTS},\II{\rightarrow},\II{\ZTS},\II{\HTS})$ 
be transition systems with the same output sets $\I{\ZTS}=\II{\ZTS}$. 
A binary relation $\rel\subset \I{\XTS}\times\II{\XTS}$ is said to be a simulation relation from $\I{\TS}$ to $\II{\TS}$ if the following three conditions are satisfied: 
\noindent\begin{enumerate} 
\item for every $\I{x}_{0}\in \I{\XTS}$, there exists $\II{x}_{0}\in \II{\XTS}$ with $(\I{x}_0,\II{x}_0)\in \rel$;
\item for every $(\I{x},\II{x})\in \rel$  we have $\I{\HTS}(\I{x})=\II{\HTS}(\II{x})$;
\item for every $(\I{x},\II{x})\in \rel$ we have that $\I{x}\I{\xrightarrow{\I{u}}}  \I{x}'$ in $\I{\TS}$ implies the existence of $\II{x}\II{\xrightarrow{\II{u}}}\II{x}'$ in $\II{\TS}$ satisfying $(\I{x}',\II{x}')\in \rel$.
\end{enumerate}
We say that $\I{\TS}$ is simulated by $\II{\TS}$, 
{or that $\II{\TS}$ simulates $\I{\TS}$},
denoted as $\I{\TS}$ \sym{similated} $\II{\TS}$, if there exists a simulation relation from $\I{\TS}$ to $\II{\TS}$.  
The transition systems $\I{\TS}$ and $\II{\TS}$ are simulation equivalent,  $\I{\TS}$ \sym{simulationequiv} $\II{\TS}$ iff $\I{\TS}\simu  \II{\TS}$  and $\II{\TS}\simu \I{\TS}.$ 
The models $\I{\TS}$ and $\II{\TS}$ are bisimilar, i.e., $\I{\TS}$ \sym{bisimilar} $\II{\TS}$, 
if there exists relation $\rel$ that is a simulation relation from $\I{\TS}$ to $\II{\TS}$ and for which $\rel^{-1}$ is also a simulation relation from $\II{\TS}$ to $\I{\TS}$.\hfill \QEDopen
\end{defn} 
 
Note that this similarity relation over the set of transition system implies a relation over the behaviour of the transition systems \cite{Tabuada2009b}, 
more precisely if $\I{\TS}\simu  \II{\TS}$ then $\mc B (\I{\TS})\subseteq \mc B (\II{\TS})$, 
and if $\I{\TS}\bsimu  \II{\TS}$ then $\mc B (\I{\TS})= \mc B (\II{\TS})$. 
% controller specifications

Approximate versions of simulation relations allow for a more robust interpretation and can be considered over metric transition systems \cite{Tabuada2009b} . 
Consider two given metric transition systems with a shared output space $\ZTS$ and a metric $\mb d$ then an $\eps$-approximate simulation relation  
The relation $\rel\subset \I{\XTS}\times \II{\XTS}$ is defined as follows (c.f. \cite{Tabuada2009b}).
\begin{defn}[Approximate Simulation Relation]\label{def:approx_sim}
Let \mbox{$\I{\TS}=(\I{\XTS},\I{\XTS}_0,\I{\UTS},\I{\rightarrow},\I{\ZTS},\I{\HTS})$}\\ and $\II{\TS}=(\II{\XTS},\II{\XTS}_0,\II{\UTS},\II{\rightarrow},\II{\ZTS},\II{\HTS})$
be transition systems with the same output space $\I{\ZTS}=\II{\ZTS}$ with metric $\mathbf{d}$. For $\eps\in\mathbb{R}^+$,  a relation $\rel\subset \I{\XTS}\times \II{\XTS}$ is said to be an $\eps$-approximate simulation relation from $ \I{\XTS}$ to $ \II{\XTS}$ if the following three conditions are satisfied: 
\noindent\begin{enumerate*} 
\item for every $\I{x}_{0}\in \I{\XTS}_0$, there exists $\II{x}_{0}\in \II{\XTS}_{0}$ with $(\I{x}_{0},\II{x}_{0})\in R$;
\item 
for every $(\I{x},\II{x})\in \rel$  we have $\mathbf{d}(\I{\HTS}(\I{x})- \II{\HTS}(\II{x}))\leq \eps$.
\item for every $(\I{x},\II{x})\in \rel$ we have that $\I{x}\I{\xrightarrow{\I{u}}}  \I{x}'$ in $\I{\TS}$ implies the existence of $\II{x}\II{\xrightarrow{\II{u}}}\II{x}'$ in $\II{\TS}$ satisfying $(\I{x}',\II{x}')\in \rel$.
\end{enumerate*}
We say that $\I{\TS}$ is \emph{approximately simulated} by $\II{\TS}$,  
  or that $\II{\TS}$ approximately simulates $\I{\TS}$, 
denoted by $\I{\TS}$\sym{similated_app} $\II{\TS}$, 
if there exists an $\eps$-approximate simulation relation from $\I{\TS}$ to $\II{\TS}$. 
The models $\I{\TS}$ and $\II{\TS}$ are approximately \emph{bisimilar}, i.e., $\I{\TS}\bsimu^\eps\II{\TS}$, 
iff there exists a relation $\rel$ that is an $\eps$-approximate simulation relation from $\I{\TS}$ to $\II{\TS}$ 
and for which $\rel^{-1}$ is an $\eps$-approximate simulation relation from $\II{\TS}$ to $\I{\TS}$. 
\end{defn}

\subsection{Formal specifications and control design}
Let us consider a specification of interest $\psi$ 
for which the desired behaviour is represented by a transition system $\TS_\psi$ \cite{Tabuada2009b}. 
%As such these similarity relation lend for 
Then a control synthesis problem for $\Sigma$ can be formulated as the search of a controller $\C$  
such that the controlled transition system, i.e.,  $\C\times\TS$ \emph{satisfies} the specification, namely 
(a.) if  $\C\times\TS\simu  \TS_{\psi}$ or (b.) if  $ \C\times\TS\bsimu  \TS_{\psi}$. The notation $\C\times\TS$ refers to the composition of the controller $\C$ with model $\TS$:  
the actions of the obtained transition system are defined by the controller $\C$, 
whereas the internal state of $\C$ is updated based on information available from $\TS$. 

If $\I{\TS}$ and $\II{\TS}$ are deterministic transition systems and $\I{\TS}\simu  \II{\TS}$, 
then for every sequence of actions for $\I{\TS}$, 
there exists a corresponding sequence for $\II{\TS}$ such that the observed behaviour is the same \cite{Girard2009}. 
%More precisely, 
Definition \ref{def:sim} suggests the refinement of a controller for $\I{\TS}$ to $\II{\TS}$ via condition 3):  for ever choice of $u_a$, picked by the controller for $\I{\TS}$,
there exists a suitable input $u_b$. 
% such that the states stay in the relation $\rel$. }
In practice this allows synthesis problems to be first solved on a simplified, and possibly finite, abstraction ($\I{\TS}$), 
before refinement over a concrete, complex {model} ($\II{\TS}$).  
%
%It is known that two \colored{system}s are in a bisimulation equivalence relation if and only if they satisfy the same CTL$^\ast$ properties, hence  both CTL and LTL properties are preserved in the bisimulation relation.
%Inherent to the satisfaction of LTL properties  is that they can expressed as in \textcolor{red}{\tiny ref katoen} via a set of allowable behaviour $\mc B_\psi$. Thus $\TS$ satisfies $\psi$ if and only if $\mc B(\TS)\subseteq\mc B_\psi$.

%More precisely, this defines the extended relation between $x_a, x_b, u_a$ as an interface $\mc F:\I{\XTS}\times \I{\UTS}\times \II{\XTS}\rightarrow \II{\UTS}$, which can be used for the construction of the refined controller $u_b$. 

\section{Problem statement}\label{prob.}
We intend
%\removed{The main objective is} 
to synthesise a certifiable output-based controller for a physical system represented by the LTI model 
\begin{align}\label{eq:LTI}
\s:\left\{\begin{array}{lll}x(t+1)&=Ax(t)+ Bu(t) & \\
y(t)&=C x(t) & \\
z(t)&=Hx(t), &
\end{array}\right.\end{align}
where $x(t)\in\Real^n$ is the state, 
initialised by $x(0)\in \X_0 \subset \Real^n$,   
the control input is $u(t)\in \Real^{m}$,  
and $y(t)\in\Real^{p}$ is the measured output available for control. 
$A,B,C $ are real matrices of appropriate dimensions.      
The signals $z(t)\in \mathbb R^q$, 
mapped from the state space via the linear map $Hx$, 
are used to define performance and properties. 
This in unlike \cite{zhang2005logic}, which defines specifications over the signals $y(t)$.  
In contrast to the measured output $y(t)$, 
the structure of which is physically specified by the sensors attached to the system, 
the choice of $H$ can be adapted to the design requirements,  
%\removed{thus} choices of $H$ \removed{may depend on specifications of interest and} 
and include $H=C$ and $H=I$ as special cases.

The LTI model $\M$ can be reinterpreted as a transition {system} characterised by a tuple $(\mathbb R^{n},\X_0,\mathbb R^m,\rightarrow,\mathbb R^q,H)$,  
with a state space $x\in\mathbb R^n$, 
a set of initial states $x(0)\in\X_0$, 
and transitions $\rightarrow:=\{x,u,x'|x'=Ax+Bu\}$. 
Additionally, $H$ assigns observation $z\in \mathbb R^{q}$ to $x\in \mathbb R^n$:  $z=H x$.
Note that this transition system has uniquely defined transitions, 
since for every state-action pair there is a unique state transition. 
 
\subsection{State-of-the-art correct-by-design controller synthesis}

Suppose that an LTI model $x(t+1)=Ax(t)+ Bu(t)$ is given, 
and that it has a finite-valued observation map that induces a partition over the observation space $\mathbb R^q$. 
Under assumptions on the controllability of the model, 
on the linear independence of the columns of its input matrix $B$, 
and on the observation map \cite{tabuada2006linear,Tabuada2009b}, 
the LTI model can be bisimulated by a finite transition system.   
Alternatively, 
under less stringent conditions it is possible to synthesise a finite approximate bisimulation of the given model \cite{Tabuada2009b,Jr2010}:  
%These finite state models are of interest, 
%since 
%\removed{as discussed in the previous subsection} 
further, 
for every controller synthesised on the finite-state abstraction there exists a refined controller for the original model, with the same closed-loop behaviour. 
%More specifically the definition of the simulation relation is constructive with respect to the refinement of state-based controller as the control \colored{system} is deterministic. 
%For a more extensive treatment of this subject please refer to \cite{tabuada2006linear,Tabuada2009b}. 

In the remainder of this work we assume that given a  model $\M$ and a model $\Sigma_\psi$ for the specification, 
both with the same output space, 
we have obtained a {controlled} 
%hybrid 
model $\bar \M_\C$, 
which is such that $\bar \M_\C\bsimu\Sigma_\psi$ \cite{Tabuada2009b}.   
$\bar \M_\C:=\C\times \M$ denotes the composition of model $\M$ with the correct-by-design controller $\C$, 
where $\C$ takes as input the state of $\M$ and returns an action to $\M$. 
This controlled model has hybrid states $(\bar x,q)$ with $\bar x\in\mathbb R^n$ and $q\in Q$, where $Q$ is a finite set. Its dynamics are defined as
\begin{align}\label{eq:symb_controlth}
 \bar \M_\C:\left\{\begin{array}{ll}\bar x(t+1)&=A\bar x(t)+B\bar u_q(\bar x(t)) 
\\ q(t+1)&=\delta(\bar x(t),q(t)), 
\end{array}\right.
\end{align}
and initialised by $(\bar x(0),q(0))\in \bigcup_{q_0\in Q_0} \left(\{q_0\}\times \X_0(q)\right) $.  
Let us remark that {the discrete states of this model follow from the states of a finite transition system, approximately bisimilar to the continuous-state model $\M$}, 
and from the discrete states of the specification model $\TS_\psi$. 
Hence the discrete state $q$ is initialised based on the specification model $\TS_\psi$ and the initial state $\bar x(0)$. 
Note that $\bar u_q(\bar x(t)) $ is a function that maps the current state to an action. 
% from which the controller is defined. 
 
\subsection{Problem statement}
Suppose that there exists a state-based, correct-by-design controller for \colored{a fully-observed LTI model, 
with closed-loop dynamics denoted by $\bar \M_\C$ as in \eqref{eq:symb_controlth}}.  
The objective of this work is to design an output-based controller, 
a controller that only requires the measured signal $y(t)$ and that can therefore be deployed on the model in \eqref{eq:LTI}.  
Additionally, 
it is required that the new controller guarantees an upper-bound on the deviation from the state-based control in \eqref{eq:symb_controlth}.  

In the following we use the notion of {\it interface function}. 
% to refine the design of a control strategy from one model to another.  
Interface functions originate from the work in \cite{Girard2009} on hierarchical control design based on (approximate) simulation relations: 
% between two deterministic models, where the 
the construction of a controller over a simplified model is \emph{refined} to a concrete model while maintaining the same guarantees over the controlled behaviour. 
%on its closed-loop behaviour. 
% 
\begin{defn}[Interface function ] \label{def:interface}\mbox{ }\\*
Let $\I{\TS}= (\I{\XTS},\I{\XTS}_0,\I{\UTS},\I{\rightarrow},\I{\ZTS},\I{\HTS})$ and  \\$\II{\TS}=(\II{\XTS},\II{\XTS}_0,\II{\UTS},\II{\rightarrow},\II{\ZTS},\II{\HTS})$
be deterministic transition systems with the same output sets $\I{\ZTS}=\II{\ZTS}$.  
A relation $\rel\subset \I{\XTS}\times \II{\XTS}$ is an $\eps$-approximate simulation relation from $\I{\XTS}$ to $\II{\XTS}$, 
and $\mathcal{F}:\I{\UTS}\times\I{\XTS}\times \II{\XTS}\rightarrow \II{\UTS}$ is its related interface, 
if the following three conditions are satisfied: 
\noindent\begin{inparaenum} 
\item for every $\I{x}_0\in \XTS_{a0}$, there exists $\II{x}_0\in \XTS_{b0}$ with $(\I{x}_0,\II{x}_0)\in \rel$;
\item for every $(\I{x},\II{x})\in \rel$, $\mathbf{d}\left(\I{\HTS}(\I{x})-\II{\HTS}(\II{x})\right)\leq \eps$;
\item for every $(\I{x},\II{x})\in \rel$ we have that $\I{x}\I{\xrightarrow{\I{u}}} \I{x}'$ in $\I{\TS}$ implies $\II{x}\II{\xrightarrow{\II{u}}}\II{x}'$  in $\II{\TS}$ with ${\II{u}}=\mathcal{F}(\I{u},\I{x},\II{x})$, satisfying $(x'_a,x'_b)\in \rel$. 
\end{inparaenum}
The feedback composition of $\I{\TS}$ and $\II{\TS}$ is denoted as $\I{\TS}\times_{\mathcal{F}}\II{\TS}$.%\\*
\mbox{ }\hfill\QEDopen \end{defn}
\smallskip 
Note that the existence of an (approximate) simulation relation implies the existence of an interface, 
i.e.,  for all $\eps$-approximately simulated and deterministic transition systems there exists at least one interface function. 

In practice Definition \ref{def:interface} entails that the dynamics corresponding to the feedback-composed models $\I{\TS}\times_{\mathcal{F}}\II{\TS}$ do not differ more than $\eps$.  
Hence, a controller composed on $\I{\TS}$ can be refined to $\II{\TS}$ via the interface $\mathcal F$, 
without affecting its closed-loop accuracy more than $\eps$. 

Let us define a specific class of interfaces denoted as {\it sensor-based interfaces}, 
which are defined exclusively based on sensor information from $\II{\TS}$, 
namely $\mathcal{F}_g: \I{\UTS}\times\I{\XTS}\times g(\II{\XTS})\rightarrow \II{\UTS}$, 
where $g$ is the sensor function. 
In the particular instance of \eqref{eq:LTI}, 
the sensor function is $g(x(t)):=Cx(t)$. 
% In the linear case of interest to us $g(x)=Cx$, see \eqref{eq:LTI}.  
These structures are of interest to us, 
as they define the set of interfaces that can be practically implemented for controller refinement on partially observable systems.  

%\removed{In the next section we extend the physical sensor information from $Cx(t)$ by adding an observer to the model dynamics:  
%as such the sensor function will include the state-estimate from the observer. }

\section{Observer-based correct-by-design controller synthesis}\label{sec:3}

In this section we propose a new design methodology for output-based controller refinement.  
We first design an observer that extends the sensors output with state estimates, see Fig. \ref{figure:MO}.  
Then as in Fig. \ref{figschematic_obs_con} we define a linear, sensor-based interface function between $\bar \M_\C$ (the state-based, correct-by-design controlled model) and the model/observer interconnection from Fig. \ref{figure:MO}.     
\subsection{Observer-based design}
\begin{figure}[t]
\centering
	\resizebox{.7\columnwidth}{!}{\begin{minipage}[b]{.8\columnwidth}\vspace{.3cm}
\tikzstyle{block} = [draw,  rectangle,
	    minimum height=1.6em, minimum width=3em]
	\tikzstyle{sum} = [draw, fill=gray!60, circle, node distance=3em]
	\tikzstyle{input} = [coordinate]
	\tikzstyle{output} = [coordinate]
	\tikzstyle{pinstyle} = [pin edge={to-,thin,black}]
 
	\begin{tikzpicture}[auto, node distance=3cm,>=latex]
	% inputs
 	\node [input, name=inputu,label=left:{$u(t)\ \ $}] {}; 
	    	% blocks
		
	\node [block, right of=inputu, minimum height =1cm,
            node distance=3em ] (system) {$\M$}; 
	\node [block, right of= system, 
            node distance=5em,yshift=.6cm,minimum height =1cm,
] (observer) {$\mathbf{O}$};	
 {};
	\node [output, node distance=9em,right of=system,yshift=-0.3cm,label=right:{$z(t)$}] (outputz) {};
     \node [output, node distance=4em,right of=observer,label=right:{$\hat x(t)$}] (xhat) {};
    	\draw[draw,->] ([xshift=-.2cm]inputu) --  (system.west);
 \draw[draw,->] (inputu) |-  ([yshift=0.3cm]observer.west);
  \draw[draw,<-] (xhat) --  (observer.east);
	\draw[draw,->]  ([yshift=0.3cm]system.east) -- node[below] {$y(t)$} ([yshift=-.3cm]observer.west);
	\draw[draw,<-] (outputz) --  ([yshift=-0.3cm]system.east);
	 
 \end{tikzpicture}\end{minipage}}
\caption{Interconnection model/observer, $\M\|\Obs(\M)$ }\label{figure:MO} 
\end{figure}
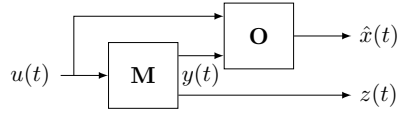

Consider a Luenberger observer denoted as $\Obs$:% 
\begin{align}\label{eq:Obs}
\hat{x}(t+1)&=A\hat{x}(t)+Bu(t)+L\left(y(t)-\hat{y}(t)\right),\\
\hat y(t)&=C \hat{x}(t),\notag
\end{align}%%\\[-  
%for which there exists
with gain matrix $L$ such that $A-LC$ is stable if $(A,C)$ is detectable \cite{franklin1998digital}.  
The observer is initialised as $\hat x(0)$, and uses the outputs from $\M$ to estimate its internal state. 
The composition of $\M$ with its observer $\Obs(\M)$ is denoted as $\M\|\Obs(\M)$ and portrayed in Fig. \ref{figure:MO}. 

Denote the sensor-based interface as%\\[- 1.4em] 
\begin{align} 
\label{eq:interface} \mathcal{F}_g(\bar u, \bar x,\hat x )=\bar u + K(\bar x -\hat x), 
\end{align}%\\[- 1.4em]
where $\bar u$ is the action selected by $\bar \M_\C$ (this role is played by $\bar u_q$ in \eqref{eq:symb_controlth}). 
For this linear interface we demand that matrix $A-BK$ is stable. 
Note that the interface is sensor-based (as defined in Section \ref{sec:2}),  
since the state estimate $\hat x$ of $x$ can be obtained from the sensor function 
%sensor signal 
of $\M\|\Obs(\M)$, thus $g(x,\hat x)=\hat x$.  
%[WHY IS THE NEXT SENTENCE RELEVANT? PLEASE ELABORATE.]
%\colored{Recall that this model can be denoted as a deterministic transition {system}, namely $\TS(\M\|\Obs(\M))=(\mathbb R^{2n},\{(\hat x_0,x_0)\},\mathbb R^m,\rightarrow,\mathbb R^n,\begin{bsmallmatrix}0&H \end{bsmallmatrix})$: 
%the state space is $(\hat x,x)\in\mathbb R^{2\times n}$, 
%the set of initial states $\{(\hat x_0,x_0)\}$, 
%and the transitions are based on \eqref{eq:LTI} and on \eqref{eq:Obs}.  
%}

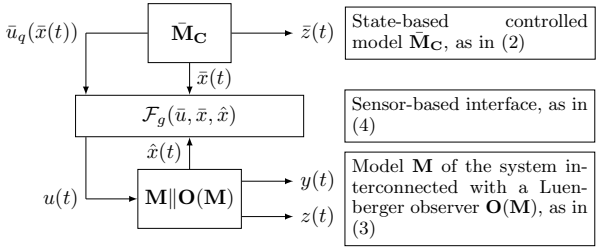
\begin{figure}[!h]
 {{\begin{minipage}[c]{\linewidth} \centering\begin{minipage}[c]{\columnwidth}
 \tikzstyle{block} = [draw,  rectangle,
	    minimum height=2em, minimum width=4em]
	\tikzstyle{sum} = [draw, fill=gray!60, circle, node distance=3em]
	\tikzstyle{input} = [coordinate]
	\tikzstyle{output} = [coordinate]
	\tikzstyle{pinstyle} = [pin edge={to-,thin,black}]
	\resizebox{.95\columnwidth}{!}{
	\begin{tikzpicture}[auto, node distance=3cm,>=latex]
	% inputs
	\begin{scope}  
	
	\node [block, minimum height =1cm,
    node distance=4em] (system) {$\Md_\C$};
      
	\node [output, node distance=5em,right of=system,
	 label=right:{$\bar z(t)$}] (outputz) {}; 
   	\node [input,left of = system,
	 name=inputu,label=left:{$\bar u_q(\bar x(t)) $}, 
	node distance=5em] {}; 
	  
    	\draw[draw ] (inputu) --  
	(system.west);
 
	\draw[draw,<- ] (outputz) --
	 (system.east); 
 \end{scope}
 \begin{scope}[yshift=-2.7cm]

	\node [block, below of=system,
	minimum width=11em, node distance=4em] 
	(interface2) {$	\mathcal{F}_g(\bar u,\bar x,\hat x)$};
 
	    \node[block,right of= outputz, minimum width=4cm, node distance=3cm](systemtext){\begin{minipage}[r]{4cm}\footnotesize State-based controlled model $\Md_\C$, as in  \eqref{eq:symb_controlth} 		\end{minipage}};
	\node [block, below of=interface2, minimum height =1cm,
    node distance=4em] (systemO) {$\mathbf{M} \|\mathbf{O}(\mathbf{M})$};
     
 	\node [output, node distance=5em,
	right of=systemO,yshift=0.3cm,
	label=right:{$y(t)$}] (outputy) {};
	\node [output, node distance=5em,right of=systemO,
	yshift=-0.3cm,label=right:{$z(t)$}] (outputz) {};
 
   	\node [input,left of = systemO, name=inputu9,label=left:{$u(t)$}, 
	node distance=5em] {}; 
%	\node [input, left of = systemO,yshift=-.3cm,name=inputw,node distance=5em, label=left:{$w(t)$}] {};  
	  \node[block,right of= interface2, minimum width=4cm, node distance=4.75cm](systemtext){\begin{minipage}[r]{4cm}\footnotesize Sensor-based interface, as in \eqref{eq:interface} 	\end{minipage}};
	
	%draw lines
    	\draw[draw,->] (inputu9) --  
	(systemO.west);
%	\draw[draw,-> ] (inputw) --  
%	([yshift=-.3cm]systemO.west);
	\draw[draw,<- ] (outputy) --  
	([yshift=.3cm]systemO.east);
	\draw[draw,<- ] (outputz) --
	 ([yshift=-0.3cm]systemO.east);
	\draw [-> ] (systemO) -- 
		node [name=y] {$\hat x(t)$}(interface2);
 	\draw [->] (system) -- node [name=y] {$\bar x(t)$}(interface2);
  	\draw [- ] (interface2.south)+(-5em,0)  -| node [name=y] { }(inputu9);
  	\draw [<-] (interface2.north)+(-5em,0) -- (inputu) ;
	 \node[block,right of= systemO, minimum width=4cm, node distance=4.75cm](systemtext){\begin{minipage}[r]{4cm} \footnotesize Model $\M$ of the system interconnected with a Luenberger observer $\Obs(\M)$, as in \eqref{eq:Obs} 
	 %for $\M$ 
	 \end{minipage}};
  \end{scope}
 \end{tikzpicture}}
 \end{minipage} \end{minipage} }}%
 \caption{Observer-based correct-by-design controller synthesis. 
 The overall interconnection is denoted as $\Mc$.
 }\label{figschematic_obs_con}
\end{figure}

The overall controlled model $\bar \M_\C \times_{\mathcal F_g} (\M\|\Obs(\M))$, 
denoted as $\Mc$, 
is the result of interfacing the two structures discussed above, 
as depicted in Fig. \ref{figschematic_obs_con}.  
This has dynamics evolving over the continuous state space $\mathbb{R}^{3n}$ as:%
\begin{align}%
\label{eq:tran3}
 \begin{array}{ll} \bar{x}(t+1) &=A \bar{x}(t)  + B \bar{{u}}_q(\bar{x}(t))\\
  \hat{x}(t+1) &=(A-LC) \hat{x}(t)  + B {u}(t)+LCx(t)\\
 x(t+1)&=Ax(t)+ B{u}(t)\\ 
 {u}(t)&=\mathcal{F}_g(\bar{{u}}_q(\bar{x}(t)), \bar x(t),\hat x(t) ) 
\end{array}
\end{align}
in combination with the discrete transitions 
\(q(t+1)=\delta(\bar x(t),q(t))\) from \eqref{eq:symb_controlth}.  
\begin{remark} 
As depicted in Fig. \ref{figschematic_obs_con}, 
we have designed an output-based controller by combining a given state-based controller with an observer.   
However, unlike classical results 
%extensions to output-based feedback 
where a state-based controller is employed over estimated states from an observer, 
in this work we have interfaced the state-based controlled model $\Md_\C$ with the model/observer interconnection $\M\|\Obs(\M)$, 
as in Fig. \ref{figure:MO}.  
This allows one to reason explicitly about the accuracy 
%and functional specifications 
of {the overall output-controlled} system, 
based on the accuracy of the sensor-based interface function. 
In special cases the proposed architecture can reduce to the classical approach.  
%however, in order to retain generality and to avoid additional conditions on the state-based controller, we do not pursue this direction in further detail.  
\hfill\QEDopen
\end{remark} 

 \begin{figure*}[!t]
\definecolor{mycolor1}{rgb}{0.83137,0.81569,0.78431}%
\definecolor{mycolor2}{rgb}{0.00000,0.44700,0.74100}%
\definecolor{mycolor3}{rgb}{0.31373,0.31373,0.31373}%
\
\begin{subfigure}[b]{.6\linewidth}	 \includegraphics[width=\linewidth]{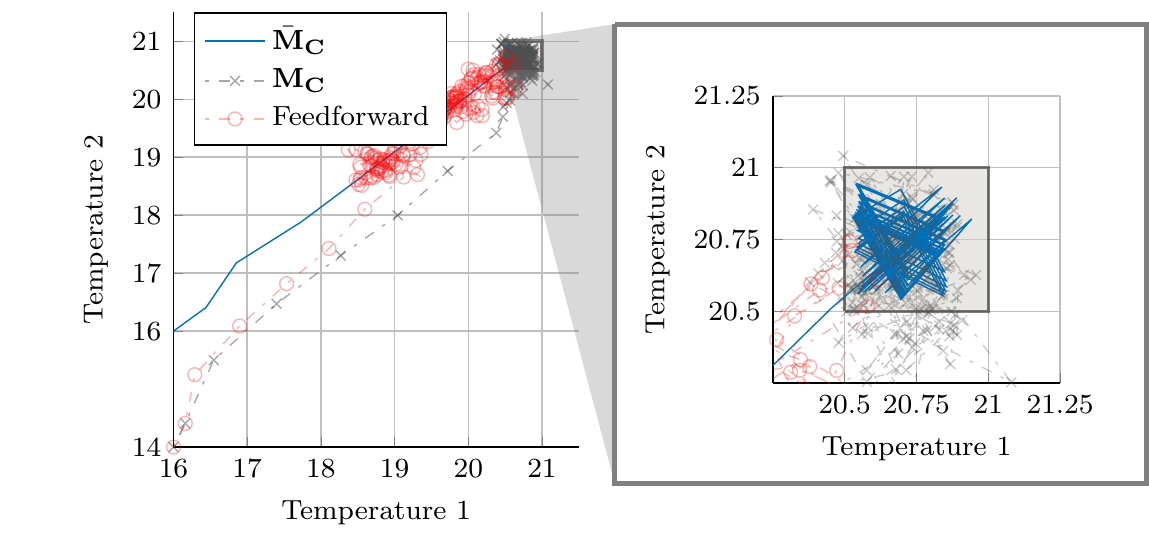}
\caption{Simulation outcomes for controlled models: $\Md_\C$ denotes state-based control of the noiseless model realisation (\cite{Jr2010}); $\M_\C$ is the output-based control of the Gaussian process model \eqref{eq:casedyn}; Feedforward denotes feedforward design using $\Md_\C$.
}\label{fig:sima}
\end{subfigure}\hspace{1cm}
\begin{subfigure}[b]{.3\linewidth}\centering
\begin{tikzpicture} \begin{axis}[
 footnotesize,
xmin=0,width=4cm,height=1.5cm,
xmax=210,
ymin=-5.5, xlabel={\footnotesize time},ylabel={\footnotesize $x(t)-\hat x(t)$},
ytick = {-5,-3,-1,1},xtick = {0,30,60,90,120,150,180,210},
ymax=1,axis x line*=bottom,
axis y line*=left,scale only axis ,baseline,trim axis left,trim axis right, xlabel near ticks,    ylabel near ticks,legend style={at={(0.75,0.45)},row sep=-1.5pt,draw=white,anchor=west,font=\footnotesize}
]
\addplot [color=blue, height=.3cm,
width=.5cm,baseline,trim axis left,trim axis right]
  table{figure4-1.tsv};
\addplot [color=black, height=.3cm,
width=.5cm,baseline,trim axis left,trim axis right]
  table{figure4-2.tsv};
  \addplot [color=purple, height=.3cm,
width=.5cm,baseline,trim axis left,trim axis right]
  table{figure4-3.tsv};
  \legend{$x_1$,$x_2$,$x_3$}
\end{axis}

\begin{axis}[yshift=-2.5cm,
 footnotesize,
xmin=0,width=4cm,height=1.5cm,
xmax=210,
ymin=-5.5, xlabel={\footnotesize time},ylabel={\footnotesize $x_3(t)$},
ytick = {-5,-3,-1,1},xtick = {0,30,60,90,120,150,180,210},
ymax=1,axis x line*=bottom,
axis y line*=left,scale only axis ,baseline,trim axis left,trim axis right, xlabel near ticks,    ylabel near ticks,
]
\addplot [color=blue, height=.3cm,
width=.5cm,baseline,trim axis left,trim axis right]
  table{figure3-1.tsv};

\end{axis}
\end{tikzpicture}%
\caption{(Upper plot) Error in state estimation for $\M_\C$; (Lower plot) Deviation from mean ambient temperature.\\}\label{fig:simb}
\end{subfigure}
\caption{Case study in smart buildings }\label{fig:sim}
\mbox{ }\\[.4em]\hrule
\end{figure*}
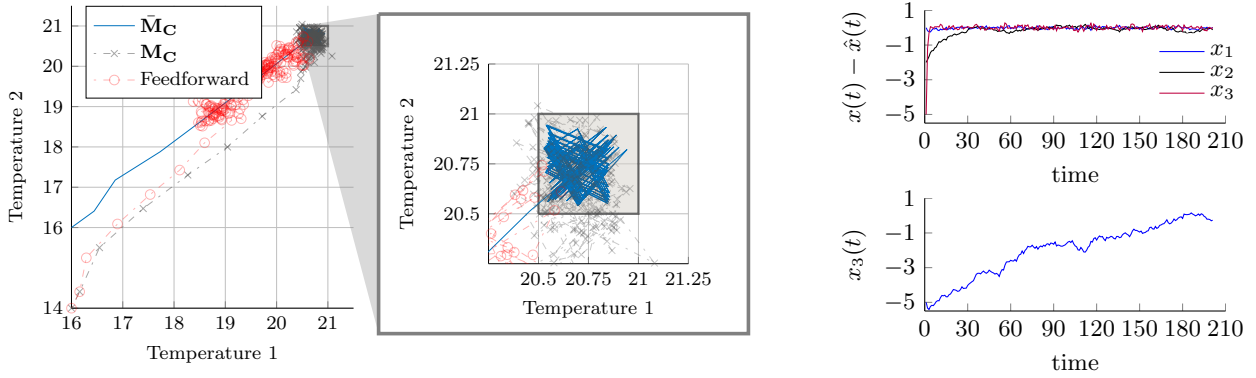

\section{Quantification of the overall accuracy}\label{sec:dev}

The controlled model $\Md_\C$, 
with traces $\bar x(t)$ as in \eqref{eq:symb_controlth}-\eqref{eq:tran3}, 
maps to the specification space as $\bar z(t)= H\bar x(t)$. 
Let a metric over this space $\mathbb R^q$ be defined as $\|\cdot\|_2$. 
Of interest is the distance between the system output $z(t)$ as in \eqref{eq:LTI} and $\bar z(t)$, 
when the system is controlled via the interconnection of Fig. \ref{figschematic_obs_con}. 
From the definition of the sensor-based interface, the following result holds. 
\begin{theorem} \label{thm:lyap} 
The function in \eqref{eq:interface} is a sensor-based interface between $\bar\M_\C$  and $\M\|\Obs(\M) $ with precision $\epsilon$, 
where% 
\begin{align} \label{eq:accuracy}\eps:=
&\textstyle\sqrt{\operatorname{trace} \big(\begin{bsmallmatrix}  H& H\end{bsmallmatrix}  Q\begin{bsmallmatrix}H&H\end{bsmallmatrix}^T\big)} \\
%\shortintertext{with}
\mbox{ with }&\begin{bsmallmatrix}
  \hat{ {x}}(0) -\bar{x}(0)\\ x(0)-\hat{x}(0)\end{bsmallmatrix} \begin{bsmallmatrix}  \hat{ {x}}(0) -\bar{x}(0)\\ x(0)-\hat{x}(0)\end{bsmallmatrix}^T-Q\preceq 0\label{eq:QX0ly}\\
 \label{eq:Qfixly}
&\begin{bsmallmatrix}A-BK&LC\\0&A-LC\end{bsmallmatrix} Q\begin{bsmallmatrix}A-BK&LC\\0&A-LC\end{bsmallmatrix}^T-\!Q\preceq0.  
 &\end{align}%
%\\[- 2em]\mbox{ } \hfill\QEDopen
\end{theorem}  
Thus the distance between $\bar{z}(t)$ and $z(t)$ is bounded by $\eps$ if there exists a $Q$  for which \eqref{eq:QX0ly} and \eqref{eq:Qfixly} are satisfied. 
A stability assumption on matrices $A-BK$ and $A-LC$ guarantees this \cite{franklin1998digital}.  
Note that since both  $\bar{x}(0)$ and $ \hat{ {x}}(0) $ are included in the design space, 
it would not make much sense to select $\bar{x}(0)\not=\hat{ {x}}(0) $ for the initialisation.
Hence, the accuracy depends on the initial states of the models only via $x(0)-\hat{x}(0)$. 
In case the initial state $x(0)$ is only known up to a set $\X_0$, 
the guarantee in Theorem \ref{thm:lyap} is required to hold over all $x(0)\in\X_0$. 
%the interface in \eqref{eq:interface} has the accuracy $\eps$. 
The proof of the theorem is given as follows.\\*
 \begin{proof} %[Theorem \ref{thm:lyap}]
The relation 
\[\begin{aligned}\textstyle &\rel:=\left\{\left(( {q},\bar{ {x}}),(\hat{ {x}},x)\right) \mid \begin{bsmallmatrix}
  \hat{ {x}}(t) -\bar{x}(t)\\ x(t)-\hat{x}(t)\end{bsmallmatrix} \begin{bsmallmatrix}\hat{ {x}}(t) -\bar{x}(t)\\ x(t)-\hat{x}(t)\end{bsmallmatrix}^T \preceq Q\right\} \end{aligned}\] 
and  \eqref{eq:interface} are a simulation relation and   interface function for the models $  \Md_\C$ and $\M\|\Obs(\M)$, since all three conditions are satisfied.
The first follows immediately from \eqref{eq:QX0ly}. The second can be shown as follows
 $z(t)-\bar z(t)=H(\hat{x}(t)-\bar{x}(t))+H(x(t)-\hat{x}(t))$,
\begin{align*}
&\|z(t)-\bar z(t)\|_2^2\leq  \begin{bsmallmatrix}
  \hat{ {x}}(t) -\bar{x}(t)\\ x(t)-\hat{x}(t)\end{bsmallmatrix}^T \begin{bsmallmatrix}H^T\\H^T \end{bsmallmatrix} \begin{bsmallmatrix}H&H \end{bsmallmatrix}\begin{bsmallmatrix}
  \hat{ {x}}(t) -\bar{x}(t)\\ x(t)-\hat{x}(t)\end{bsmallmatrix} \\
  &\qquad =\operatorname{trace}\left(\begin{bsmallmatrix}H&H \end{bsmallmatrix}\begin{bsmallmatrix}
  \hat{ {x}}(t) -\bar{x}(t)\\ x(t)-\hat{x}(t)\end{bsmallmatrix} \begin{bsmallmatrix}\hat{ {x}}(t) -\bar{x}(t)\\ x(t)-\hat{x}(t)\end{bsmallmatrix}^T \begin{bsmallmatrix}H^T\\H^T \end{bsmallmatrix} \right) 
\shortintertext{ if $\bar{ {x}}, \hat{ {x}},x \in \rel$ then }
&\|z(t)-\bar z(t)\|_2^2\leq \operatorname{trace}\big(\begin{bsmallmatrix}H&H \end{bsmallmatrix}Q \begin{bsmallmatrix}H^T\\H^T \end{bsmallmatrix} \big)=\eps^2  
\shortintertext{and $\|z(t)-\bar z(t)\|_2\leq \eps $.
The third condition follows, suppose that $\bar{ {x}}(t), \hat{ {x}}(t),x(t) \in \rel$ then }
&\begin{bsmallmatrix}\hat{ {x}}(t+1) -\bar{x}(t+1)\\ x(t+1)-\hat{x}(t+1)\end{bsmallmatrix}=\begin{bsmallmatrix}A-BK&LC\\ 0& A-LC\end{bsmallmatrix} \begin{bsmallmatrix}\hat{ {x}}(t) -\bar{x}(t)\\ x(t)-\hat{x}(t)\end{bsmallmatrix} 
\shortintertext{thus $\bar{ {x}}(t+1), \hat{ {x}}(t+1),x(t+1) \in \rel$ if}
&\begin{bsmallmatrix}A-BK&LC\\0&A-LC\end{bsmallmatrix} Q\begin{bsmallmatrix}A-BK&LC\\0&A-LC\end{bsmallmatrix}^T\preceq Q \end{align*} which holds due to \eqref{eq:Qfixly}.
%
%Apply a congruence transform with $\Delta_x(t)=\hat{x}(t)-\bar{x}(t)$ and ${e}(t)=x(t)-\hat{x}(t)$ :
%{\small \begin{align*} %\label{eq:tran3}
% \begin{array}{lll} 
% \bar{x}(t+1) &=A \bar{x}(t)  + B \bar{ {u}}_q(\bar{x}(t))\\
%\Delta_x(t+1) &=(A-BK) \Delta_x(t) +LC {e}(t)\\
% { e}(t+1) &= (A-LC) {e}(t) \\
%     { z}(t)&= H\bar{x}(t)+H\Delta_{x}(t)+H {e}(t).&  \end{array} 
%\end{align*}}%  . 
%Then $z(t)-\bar z(t)=H\Delta_{x}(t)+H{e}(t)$
\end{proof}\bigskip
   
\section{Stochastic disturbances: robustness}\label{sec:4}
%%%%%%%%%%%%%%%%%%%%%%%%%%%%%%%%%%%%%%%%%%%%%%%%%%%%%%%%%%%%%%%%%%%%%%%%%%%%%%%%  
 
We extend the previous results supposing that the physical system $\M$ is disturbed by stochastic noise.  
More precisely, 
state transitions are affected by additive noise $\mb w_1(t)$ with realisations $w_1(t)\sim\mb w_1(t)$ taking values in $\mathbb R^{d_1}$, 
whereas sensor measurements are disturbed by noise sources $\mb w_2(t)$, 
with realisations $w_2(t)\sim\mb w_2(t)$ in $\mathbb R^{d_2}$. 
(We denote random variables $\mb x$ as bold faced, in contrast to their realisations $x\sim\mb  x$.) 
Each of the noise sources is supposed to be independent and identically distributed over time, 
with zero mean and unit variance. 
This assumption holds for a typical Gaussian process noise with distribution $w_1(t)\sim\mc N(0,I_{d_1\times d_1})$. 
The resulting stochastic model is  %\\[- 1em]
\begin{align}\label{eq:LTIst}
\M:\left\{\begin{array}{lll}x(t+1)&=Ax(t)+ B u(t)+F  w_1(t)& \\
  y(t)&=C x(t) +E  w_2(t)&\\
  z(t)&=H  x(t), &
\end{array}\right.\end{align} %\\[- 1em]
where the matrices $ F ,E,$ are again real-valued matrices of appropriate dimensions.   
The model is initialised as $\mb x(0)\sim \mathcal{N}(x_0,P_0)$.  

With reference to the previous section, 
the control design strategy is as follows:  
\begin{enumerate}[A.]
\item Let $\Md$ be a noiseless version of $\M$ in \eqref{eq:LTIst}, 
and $\bar\M_\C$ be the composition of $\Md$ with its correct-by-design controller;   
\item Design a state observer  $\Obs(\M)$ for $\M$;% $K\in \mathbb R^{1\times n}$, $L\in\mathbb R^{n\times 1}$, $Q\in \mathbb R^{2n\times 2n}$ and $\eps\in \mathbb R^+$ such that \[
 %\begin{aligned}Q&=\begin{bmatrix}A-BK&LC\\0&A-LC\end{bmatrix} Q\begin{bmatrix}A-BK&LC\\0&A-LC\end{bmatrix}^T\\&\quad +\begin{bmatrix}LE\\ F-LE  \end{bmatrix}\begin{bmatrix}LE\\ F-LE  \end{bmatrix}^T  &\end{aligned}\] and\( \operatorname{trace} \left(\begin{bmatrix}  H& H\end{bmatrix} Q\begin{bmatrix}H^T\\H^T\end{bmatrix}\right) \leq\eps^2.\)
\item Design a linear interface function $\mathcal{F}_g$ stabilising ${A-BK}$;
\item Implement the control structure in Fig. \ref{figschematic_obs_con}, 
and denote the resulting controlled stochastic {model} as $\M_\C:=\Md_\C\times_{\mc F_g}\left(\M\|\Obs(\M)\right)$. 
\end{enumerate}
The initial conditions for $\M_\C$, 
namely $\bar x(0), \hat x(0)$,  
are selected as part of the control design problem:  
as discussed earlier, 
we pick $\bar x(0)=\hat x(0)$. 
Further, 
let $q(0)$ be any discrete state such that $(\bar x(0),q(0))\in \bigcup_{q_0\in Q_0} \left(\{q_0\}\times \X_0(q)\right) $.
%, since any other choice would not be rational. 
 
In order to analyse the behaviour of the controlled stochastic {model} $\M_\C$ with respect to a metric of interest, 
let us embed $\M_\C$ into the formalism of deterministic transition systems (cf. Definition \ref{def:trans}) as in \cite{Zamani}. 
The 
%controlled stochastic 
{model} 
%$\Mc$ 
can be represented as a symbolic transition {system} 
%(cf. Definition \ref{def:trans}) 
$\Sigma^\ast(\Mc)$, 
with states encompassing random variables $\xv_\C(t)$ representing the distribution of $x_\C(t)\sim \xv_\C(t)$, with $x_\C(t)\in \mathbb R^{3n}$ as in \eqref{eq:tran3}. 
Consider the metric output space $\ZTS$, 
to which the states are mapped as \(\mathbf{z}_\C(t)= H\mathbf{x}_\C(t). \)
%Thus the output $\mathbf{z}_\C(t)$ of this symbolic \colto{model} is also a random variable, for which $z_\C(t)\sim \mathbf{z}_\C(t)$.  The metric on the output is defined as 
Further consider the metric $\mathbf d^\ast(\mathbf{z}_1-\mathbf{z}_2)=\mathbb E(\| \mathbf{z}_1-\mathbf{z}_2 \|_2 )$, with $\|\cdot\|_2$ the Euclidean norm. 
Denote the set of all transition systems with the metric output space $\ZTS$ as $\mathcal{T}^\ast$.

Both the specification {model} $\TS_\psi$ and the correct-by-design controlled {model} $\Md_\C$ can be trivially embedded in $\mathcal{T}^\ast$ via singleton distributions: 
we denote the corresponding symbolic transition systems as  $\Sigma^\ast_\psi$ and $\TS^\ast(\Md_\C)$, respectively. 
We obtain: 
 \begin{theorem}\label{thm:h2} Transition {system} $\TS^\ast(\Mc)$ is approximately bisimulated by $\TS^\ast(\Md_\C)$ with precision $\epsilon$ obtained as 
\begin{align} \label{eq:prech2}&\allowdisplaybreaks
\textstyle\eps:=\textstyle\sqrt{\operatorname{trace} (\begin{bsmallmatrix}  H& H\end{bsmallmatrix}  Q\begin{bsmallmatrix}  H& H\end{bsmallmatrix} ^T)}\\  &\textmd{with } \begin{bsmallmatrix} 0&0\\ 0& (x_0-\hat x(0))(x_0-\hat x(0))^T\end{bsmallmatrix}+\begin{bsmallmatrix} 0&0\\ 0& P_0\end{bsmallmatrix}-Q\preceq 0\label{eq:QP0}\\
 \label{eq:Qfix}
&\begin{bsmallmatrix}A-BK&LC\\0&A-LC\end{bsmallmatrix} Q\begin{bsmallmatrix}A-BK&LC\\0&A-LC\end{bsmallmatrix}^T\notag\\&+\begin{bsmallmatrix}LEE^TL^T& -LEE^TL^T\\ -LEE^TL^T& FF^T+LEE^TL^T  \end{bsmallmatrix}  -Q\preceq0\, .\ \QEDopen 
 \end{align}%
 \end{theorem}
%
 %Note that  { \scriptsize  \begin{align*}  &\mb{E}\left[\begin{bmatrix} (\hat{\xv}(t+1)-\bar{\xv}(t+1))\\ (\xv(t+1)-\hat{\xv}(t+1))\end{bmatrix} \begin{bmatrix} (\hat{\xv}(t+1)-\bar{\xv}(t+1))\\ (\xv(t+1)-\hat{\xv}(t+1))\end{bmatrix} ^T \right]\\&=   \begin{bmatrix}A-BK &LC\\0&A-LC\end{bmatrix} \mb{E}\left[\begin{bmatrix} (\hat{\xv}(t)-\bar{\xv}(t))\\ (\xv(t)-\hat{\xv}(t))\end{bmatrix} \begin{bmatrix} (\hat{\xv}(t)-\bar{\xv}(t))\\ (\xv(t)-\hat{\xv}(t))\end{bmatrix} ^T \right]\\&\quad\times\begin{bmatrix}A-BK &LC\\0&A-LC\end{bmatrix}^T\notag + \begin{bmatrix}  LE\\ F-LE\end{bmatrix}\begin{bmatrix}  LE\\ F-LE\end{bmatrix}^T\notag \end{align*}} 
As a consequence\footnote{
Note that we have trivially assumed that this (bi-)~simulation relation between the transition  {system} $\TS(\Md_\C)$ and $\TS_\psi$ is maintained when embedding them in $\mathcal{T}^\ast$ via Dirac distributions \cite{Zamani}. } of Theorem \ref{thm:h2} it follows that if 
$\TS^\ast(\Md_\C)\simu\TS^\ast_\psi$,  then $\TS^\ast(\M_\C)\simu^{\eps}\TS^\ast_\psi$, 
and if $\TS^\ast(\Md_\C)\bsimu\TS^\ast_\psi$,  then $\TS^\ast(\M_\C)\bsimu^{\eps}\TS^\ast_\psi$. 
%The proof of Theorem \ref{thm:h2} follows in the Appendix.
Finally note that \eqref{eq:Qfix} is known to admit positive matrices $Q$ for which $\eps$ is finite if $A-BK$ and $A-LC$ are both stable matrices \cite{franklin1998digital}.%This follows from the computation of $H_2$ norm or optimal linear quadratic Gaussian control. 

  \begin{proof} 
 % [Theorem \ref{thm:h2}] 
 The composition of $\Md_\C$ with $\mathbf{M} \|\mathbf{O}(\mathbf{M})$  over the interface \eqref{eq:interface} gives the continuous dynamics of as $\TS^\ast(\Mc)$%
\begin{align*} %\label{eq:tran3}
 \begin{array}{lll} 
 \bar{\xv}(t+1) &=A \bar{\xv}(t)  + B \bar{\mb{u}}_q(\bar{\xv}(t))\\
  \hat{\xv}(t+1) &=(A-LC) \hat{\xv}(t)  + B \mb{u}(t)+LC\xv(t)+ LE \mb{w}(t)\\
 \xv(t+1)&=A\xv(t)+ B\mb{u}(t)+F \mb{w}(t)\\ 
 \mb{u}(t)&=\bar{\mb{u}}_q(\bar{\xv}(t)) +K(\bar{\xv}(t)-\hat{\xv}(t)) 
\end{array} 
\end{align*}%
with output $
     \mb{z}(t)= H\xv(t) $.  Consider the relation $\mathcal{R}$ defined as 
\[\begin{aligned}
&\mathcal{R}:=\left\{\left((\mb{q}',\bar{\mathbf{x}}'),(\mb{q},\bar{\mathbf{x}},\hat{\mathbf{x}},\xv)\right) \mid \bar{\mathbf{x}}'=\bar{\mathbf{x}}, \mb{q}'=\mb{q}, \right.\\&\quad\left.
\wedge \mb{E}\left[[ (\hat{\mathbf{x}} -\bar{\xv})^T\ (\xv-\hat{\xv})^T ]^T [ (\hat{\mathbf{x}} -\bar{\xv})^T\ (\xv-\hat{\xv})^T ] \right]\preceq Q\right\}\end{aligned}\]%
 where $\bar{\xv}'$ is the continuous state of $\TS^\ast(\Md_\C)$ and $\xv_\C:=\begin{bsmallmatrix}\bar{\mathbf{x}}^T\ \hat{\mathbf{x}}^T\ \xv^T\end{bsmallmatrix}^T$ the continuous state of $\TS^\ast(\Mc)$. The outputs $\mb{z}'$ and $\mb{z}$ are similarly defined. 
For future reference note that applying the congruence transform with $\Delta_\xv(t)=\hat{\xv}(t)-\bar{\xv}(t)$ and $\mb{e}(t)=\xv(t)-\hat{\xv}(t)$ gives :  \begin{align*} %\label{eq:tran3}
 \begin{array}{lll} 
 \bar{\xv}(t+1) &=A \bar{\xv}(t)  + B \bar{\mb{u}}_q(\bar{\xv}(t))\\
\Delta_\xv(t+1) &=(A-BK) \Delta_\xv(t) +LC\mb{e}(t)+LE \mb{w}(t) \\
\mb{ e}(t+1) &= (A-LC)\mb{e}(t)  +( F-LE)\mb{w}(t)\\ 
    \mb{ z}(t)&= H\bar{\xv}(t)+H\Delta_{\xv}(t)+H \mb{e}(t).&  \end{array} 
\end{align*}%  . 
Firstly  condition 1)  for an approximate bisimulation holds : $\bar{\mathbf{x}}'(0)=\bar{\mathbf{x}}(0)$,   $ \hat{\mathbf{x}}(0) -\bar{\xv}(0)=0$ and  $\mb E\left[\left(\xv(0) -\hat{\mathbf{x}}(0) \right)\left(\xv(0) -\hat{\mathbf{x}}(0) \right)^T\right]=(x_0-\hat x(0))(x_0-\hat x(0))^T+P_0$. Therefore based on \eqref{eq:QP0} it follows that the first condition holds.
Secondly for all 
$  ((\mb q',\bar{\mathbf{x}}'),(\mb q, \bar{\mathbf{x}},\hat{\mathbf{x}},\xv))\in \mathcal{R}:
$ the metric $\mb{E}\left[\|\mb{z}'-\mb{z}\|_2\right]$ can be written as
 \[\begin{aligned}
\mb{E}\left[\|H\bar{\xv}'-H\xv\|_2\right] 
 \leq \sqrt{\mb{E}\left[(H\bar{\xv}'-H\xv)^T (H\bar{\xv}'-H\xv) \right]}\\
 =\sqrt{\operatorname{trace} \mb{E}\left[H(\bar{\xv}'-\xv) (\bar{\xv}'-\xv)^T H^T \right]}.
 \end{aligned}\]  Note that  $(\bar{\xv}' -\xv)=((\bar{\xv}'-\bar{\xv}) -(\hat{\xv}-\bar{\xv})-(\xv-\hat{\xv})) $, and $\bar{\xv}'-\bar{\xv} =0$ (due to the relation $\rel$), then the metric is bounded from above by
 \[\begin{aligned}
 \operatorname{trace}\left(  \begin{bsmallmatrix} H^T\\ H^T\end{bsmallmatrix}^T \!\!\mb{E}\left[\begin{bsmallmatrix} (\hat{\xv}-\bar{\xv})\\ (\xv-\hat{\xv})\end{bsmallmatrix} \begin{bsmallmatrix} (\hat{\xv}-\bar{\xv})\\ (\xv-\hat{\xv})\end{bsmallmatrix} ^T \right]\begin{bsmallmatrix} H^T\\H^T\end{bsmallmatrix}\right)^{\frac{1}{2}} 
\\\leq  \operatorname{trace}\left( \begin{bsmallmatrix} H^T\\ H^T\end{bsmallmatrix}^TQ\begin{bsmallmatrix} H^T\\ H^T\end{bsmallmatrix}\right)^{\frac{1}{2}}.   
 \end{aligned}
 \]%
For the third condition we have to prove invariance of $\rel$. More specifically if  $ ((\mb q'(t),\bar{\mathbf{x}}'(t)),(\mb q(t), \bar{\mathbf{x}}(t),\hat{\mathbf{x}}(t),\xv(t)))\in \mathcal{R}$ than for every transitions of $\Md_\C$: $(\mb q'(t),\bar{\mathbf{x}}'(t))\rightarrow (\mb q'(t+1),\bar{\mathbf{x}}'(t+1))$ there exists a transient in $\M_\C$ for which $ ((\mb q'(t+1),\bar{\mathbf{x}}'(t+1)),(\mb q(t+1), \bar{\mathbf{x}}(t+1),\hat{\mathbf{x}}(t+1),\xv(t+1)))\in \mathcal{R}$. 

Since $\M_\C$ is composed of $\Md_\C\times_{\mc F_g}(\M \|\Obs(\M))$ we know that for every transition $(\mb q'(t),\bar{\mathbf{x}}'(t))\rightarrow (\mb q'(t+1),\bar{\mathbf{x}}'(t+1))$ in $\Md_\C$ there is an equivalent transition $(\mb q(t),\bar{\mathbf{x}}(t))\rightarrow (\mb q(t+1),\bar{\mathbf{x}}(t+1))$ in $\M_\C$. And the congruence transformed dynamics are such that 
\begin{align*}  &\mb{E}\left[\begin{bsmallmatrix} (\hat{\xv}(t+1)-\bar{\xv}(t+1))\\ (\xv(t+1)-\hat{\xv}(t+1))\end{bsmallmatrix} \begin{bsmallmatrix} (\hat{\xv}(t+1)-\bar{\xv}(t+1))\\ (\xv(t+1)-\hat{\xv}(t+1))\end{bsmallmatrix} ^T \right]\\&=   \begin{bsmallmatrix}A-BK &LC\\0&A-LC\end{bsmallmatrix} \mb{E}\left[\begin{bsmallmatrix} (\hat{\xv}(t)-\bar{\xv}(t))\\ (\xv(t)-\hat{\xv}(t))\end{bsmallmatrix} \begin{bsmallmatrix} (\hat{\xv}(t)-\bar{\xv}(t))\\ (\xv(t)-\hat{\xv}(t))\end{bsmallmatrix} ^T \right]\\&\quad \times\begin{bsmallmatrix}A-BK &LC\\0&A-LC\end{bsmallmatrix}^T\notag + \begin{bsmallmatrix}LEE^TL^T& -LEE^TL^T\\ -LEE^TL^T& FF^T+LEE^TL^T  \end{bsmallmatrix}
 \preceq Q
 \end{align*}%  
because \eqref{eq:Qfix}  holds.  Therefore every transition of $\Md_\C$ can be mimicked by $\M_\C$. The proof that every transition in $\M_\C$ has an equivalent transition in $\Md_\C$ goes along the same lines.  \end{proof}

\subsection{Selection of the matrix gains $L$ and $K$}\label{sec:SelectLK}
Thus far we have assumed that $L$ and $K$ are chosen so that they stabilise $A-LC$ and $A-BK$. 
It is known that, as long as the model is detectable and stabilisable, 
these gains exist \cite{franklin1998digital}. 
A constructive approach to obtain $L,K$ in a semi-optimal manner follows from Theorem \ref{thm:h2}. Omitting the initialisation, 
the computation of the precision level defined in \eqref{eq:prech2} together with \eqref{eq:Qfix} for given $L$ and $K$ is equivalent to  
\(\eps=\lim_{t\rightarrow \infty}\sqrt{\mathbf{E} \|\Delta z(t)\|_2^2  }\) for %\\[- 1.2em]
\begin{align}
\Delta x(t+1) &=\begin{bsmallmatrix}A-BK&LC\\0&A-LC\end{bsmallmatrix}  \Delta x(t)\\&\qquad+ \begin{bsmallmatrix}0&LE\\ F&-LE  \end{bsmallmatrix}\begin{bsmallmatrix} w_1(t)\\w_2(t)\end{bsmallmatrix}\notag \\
\Delta z(t)&=Hx(t), 
\end{align} 
for given white noise sequences $w_1(t),w_2(t)$. 
As such the optimisation problem leading to $L$ and $K$ has been recast in the familiar LQG stochastic control problem \cite{Witsenhausen1971} for which it is known that the optimal observer gain $L$ and the optimal state-feedback gain $K$ can be computed separately. 
The optimal observer gain with respect to the LQG problem is the Kalman filter gain,  \( L^\ast=\left(A PC^T\right)\left(CPC^T+EE^T\right)^{-1}\) %
 s.t. 
 \(P=APA^T-(APC^T)(CPC^T+EE^T)^{-1}(CPA^T)+FF^T.\)
On the other hand, 
the optimal state-feedback gain $K$ solves a quadratic control problem, 
that is 
% $K^\ast$, 
\(  K^\ast=(B^TSB)^{-1}B^TSA\) s.t.  \[S  =A^TSA-A^TS B(B^TS B)^{-1}B^T SA+H^T H.\] 
In the next case study this will be computed via the generalised eigenproblem algorithm \cite{arnold1984generalized} implemented in MATLAB. 
% \cite{MATLAB:2015a}. 
Note that since there is no trade-off between the state error and the magnitude of the control gain, 
the state feedback gain will push the control to deadbeat control \cite{franklin1998digital}:  
this behaviour can be easily remedied by extending the observation space $H$ with $D_H$, 
such that the extended performance signal becomes $\cramped{z_e(t)=\begin{bmatrix}z^T(t)& z_u^T(t)\end{bmatrix}^T}$, 
with $z_u(t)=D_H  K x(t)$, 
or equivalently with $z_u(t)=D_H  \left(u(t)-\bar u(t)\right)$.  
% 
% Substitute $Q$ by $M=\eps Q^{-1}$,  and introduce the slack variable  
% {\footnotesize  $Z:= \begin{bmatrix}  C& C\end{bmatrix} Q\begin{bmatrix}C^T\\C^T\end{bmatrix}+\gamma I$ } for which there exists $\gamma>0$ such that $\operatorname{trace}(Z)<\eps.$
% 
 
 \section{Case study in Smart Buildings} \label{sec:5}
% 1. Describe case study
We are interested in the advanced energy management of an office building.
As a motivation for output-based controllers, 
consider a building that is divided in two connected zones, 
each with a radiator regulating the heat in each zone via the controlled boiler water temperature \cite{Holub2013}.  
Due to a sensor fault in the second zone, only the temperature in the first zone and the ambient (outside) temperatures are measured. 
The temperature fluctuations in the two zones and the ambient temperature are modelled via $\M$ as 
%a Gaussian process 
\cite{Holub2013} 
{\begin{align}
 x(t+1)&=\Xi x(t)+ \Gamma  u(t) + Fw_1(t)\label{eq:casedyn}\\
y(t)&=\begin{bsmallmatrix}1&0&0\\0&0&1\end{bsmallmatrix}x(t)+ E w_2(t),\quad
z(t)=\begin{bsmallmatrix}1&0&0\\0&1&0\end{bsmallmatrix}x(t), 
\shortintertext{with stable dynamics}
\Xi&=\begin{bsmallmatrix}
    0.8725  &  0.0625 &   0.0375\\
    0.0625  &  0.8775 &   0.0250\\
         0  &       0 &   0.9900\end{bsmallmatrix}\notag%
,\quad \Gamma =\begin{bsmallmatrix}
    0.0650&         0\\
         0&    0.0600\\
         0&         0
\end{bsmallmatrix} ,
\end{align}}%
where $x_{1,2}(t)$ are the temperatures in zone 1 and 2, respectively; 
$x_{3}(t)$ is the deviation of the ambient temperature from its mean; 
and $u(t)\in\mathbb{R}^2$ is the control input.  
Note that since $\Xi$ is stable, it follows that $(\Xi,\Gamma)$ is stabilisable and $(\Xi,\begin{bsmallmatrix}1&0&0\\0&0&1\end{bsmallmatrix})$ is detectable.  
The state variables are initiated as $x(0)=[16\ 14\ -5]^T$. 
The constants in matrix $\Xi$ are selected to represent the heat exchange rate between the individual zones and the heat loss rate of each zone to the ambient;  
those in $ \Gamma$ represent the rate of heat supplied by the radiators to the two zones, respectively.  
The disturbances are modelled as independent and identically distributed standard normal distributions 
$w_{1,2}(t)$, rescaled by 
{ \[ F=\begin{bsmallmatrix} .05&    -.02  &       0\\
   -.02   &  .05    &      0\\
         0    &     0 &   0.1 \end{bsmallmatrix}\textmd{ \normalsize and }E=\begin{bsmallmatrix}.05&0\\0&.05\end{bsmallmatrix}.
\]}%
The upper block in $F$  represents random heat transfers, 
caused for example by people moving within and between zones, 
whereas the lower, right-diagonal element represents the stochastic nature of the fluctuation in the outside temperature. 
The values in $E$ define the standard deviation of the additive disturbance on the temperature sensors in the first zone and in the ambient. 
$y(t)$ is the stochastic signal that can be measured, whereas the specification is defined over $z(t)$ (zone temperatures).

The objective is to design an output-based, correct-by-design controller, 
such that the temperature trajectories $z(t)=(x_1(t),x_2(t))$ eventually both take values in the interval $\left[20.5,\ 21\right]^2$, 
and remain within this interval thereafter.\footnote{This property can be formally expressed as an ``eventually always'' specification in LTL.} 
% namely $\eventually\always$.}  
The controller is initialised with $\hat x(0)=\left[16\ 16\ 0\right]^T$: 
this deviation from $x(0)$ is selected to model a realistic situation occurring after a sensor failure in zone 2 is discovered. 

The dynamics of the noiseless model $\Md$ are solely governed over the first two states, 
where the correct-by-design controller for the given specification is designed. 
We synthesise $\Md_\C$ by \textmd{PESSOA} \cite{Jr2010}, 
where the discrete-time dynamics are further discretised over state and action spaces:  
we have selected 
%as abstraction parameters 
a state quantisation of $.05$ over the range $\left[15, 25\right]^2$, and an input quantisation of $.05$ over $\left[10, 30\right]^2$.   
Fig. \ref{fig:sima} displays (continuous blue line) the state trajectory of the obtained correct-by-design system $\Md_\C$: %, 
%implemented on the noiseless model $\Md$ (continuous blue line): 
it can be observed that the controller regulates the  model to eventually remain within the target region. 

Next, we are interested in extending the designed controller to  the concrete (noisy) model of the system based on noisy output measurements of the first zone and of the ambient. 
As a first attempt we implement the controller based on a feedforward architecture, where $\mc{F}_{ff}:=\bar u(t)$. This is what we would obtain applying the results in \cite{Zamani}. It can be observed in Fig. \ref{fig:sima} (circled red realisation) that a trajectory $(x_1(t),x_2(t))$ in $\Md_\C\times_{\mc{F}_{ff}}\M$ deviates substantially from the desired temperature range. In Table \ref{TabI} the accuracy of this feedforward interface is given.
As a second 
%, improved 
design, we implement the structure in Fig. \ref{figschematic_obs_con}, 
where the gains $K,L$, 
as detailed in Subsection \ref{sec:SelectLK}, 
are selected as the optimal LQ and Kalman gains, respectively. 
The resulting design values are
\begin{equation*}
L=\begin{bsmallmatrix}    0.5201 &   0.0333\\
   -0.2239 &   0.0262\\
    0.0022  &  0.8196\\
\end{bsmallmatrix}\textmd{ and }\, K = \begin{bsmallmatrix}   13.4231&    0.9615 &   0.5769\\
    1.0417  & 14.6250  &  0.4167
\end{bsmallmatrix}.
\end{equation*} 
A trajectory (crossed grey line in Fig. \ref{fig:sima}) realised from $\M_\C=\Md_\C\times_{\mc{F}_g} (\M\|\Obs(\M))$ and based on the previous noise realisation ends up close to the desired temperature range.\begin{table}[!t]
\centering
\caption{\it Error Bounds -- Accuracy of the controlled systems based on the interface. 
An initialisation is given by $\eps_{x_0}$, for the perfect initialisation, or for $t\rightarrow\infty$ the system the accuracy is given as $\eps_{\infty}$. The estimates  $\cramped{\hat\eps_{x_0,100}}$ and $\eps_\infty$ are computed as $\cramped[\scriptstyle]{\sqrt{ \hat{\mathbf{E}}_{1:100}\|z(t)-\bar z(t)\|_2^2}}$ and $\cramped[\scriptstyle]{\sqrt{ \hat{\mathbf{E}}_{10^2: 4\times 10^3}\|z(t)-\bar z(t)\|_2^2}}$ respectively, with the empirical mean computed as $\cramped[\scriptstyle]{\mathbf{E}_{i:j }x=\frac{1}{j-i}\sum_{k=i}^jx(k)} $.}\label{TabI}
\begin{tabular}{|l|ll|ll|}\hline
&$\eps_{x_0}$& $\eps_\infty$&$\hat\eps_{x_0,100}$& $\hat\eps_{\infty}$ \\\hline\hline
$\Md_\C\times_{\mc{F}_{ff}}\M$& 3.9618& 0.4890&1.9961 &0.4845\\
$\M_\C$&2.1194&0.1284& 0.5184& 0.1240\\ \hline
\end{tabular} \end{table} This substantial improvement with respect to the feedforward interface is also quantified in Table \ref{TabI}\,.
Fig. \ref{fig:simb} displays the error of the state estimation $x(t)-\hat x(t)$ of $\M_\C$ (upper plot):  
it can be observed that the estimated state converges to the exact state. 
The lower plot in Fig. \ref{fig:simb} provides a simulation of the deviation of the ambient temperature from its mean. 
 
\section{Conclusions and future work}
In this work we have shown that correct-by-design controllers can be extended to work on {stochastic} partially-observable LTI systems, 
as long as the LTI system is detectable and stabilisable.
Future work will concern extensions to non-linear dynamics and the development of tailored notions of probabilistic approximations. 

%%%%%%%%%%%%%%%%%%%%%%%%%%%%%%%%%%%%%%%%%%%%%%%%%%%%%%%%%%%%%%%%%%%%%%%%%%%%%%%%

\bibliographystyle{abbrv}
 
\bibliography{library2}

\begin{thebibliography}{10}

\bibitem{Abate2011}
A.~Abate.
\newblock Approximation metrics based on probabilistic bisimulations for
  general state-space markov processes: A survey.
\newblock {\em Electronic Notes in Theoretical Computer Science}, 297:3 -- 25,
  2013.
\newblock Proceedings of the first workshop on Hybrid Autonomous Systems.

\bibitem{arnold1984generalized}
W.~F. {Arnold III} and A.~J. Laub.
\newblock Generalized eigenproblem algorithms and software for algebraic
  {Riccati} equations.
\newblock {\em Proceedings of the IEEE}, 72(12):1746--1754, 1984.

\bibitem{Chatterjee2014}
K.~Chatterjee, M.~Chmelik, R.~Gupta, and A.~Kanodia.
\newblock Qualitative analysis of {POMDP}s with temporal logic specifications
  for robotics applications.
\newblock {\em CoRR}, abs/1409.3360, 2014.

\bibitem{Clarke2008}
E.~M. Clarke.
\newblock The birth of model checking.
\newblock In {\em 25 Years of Model Checking}, pages 1--26. Springer, 2008.

\bibitem{franklin1998digital}
G.~F. Franklin, J.~D. Powell, and M.~L. Workman.
\newblock {\em Digital control of dynamic systems}.
\newblock Addison-Wesley Menlo Park, second edition, 1990.

\bibitem{Girard2009}
A.~Girard and G.~J. Pappas.
\newblock {Hierarchical control system design using approximate simulation}.
\newblock {\em Automatica}, 45(2):566--571, 2009.

\bibitem{Giro2012}
S.~Giro and M.~N. Rabe.
\newblock Verification of partial-information probabilistic systems using
  counterexample-guided refinements.
\newblock In {\em Proc. on Automated Technology for Verification and Analysis},
  LNCS, pages 333--348. Springer, 2012.

\bibitem{Holub2013}
O.~Holub and K.~Macek.
\newblock {HVAC simulation model for advanced diagnostics}.
\newblock In {\em Symp. Intelligent Signal Processing}, pages 93--96. IEEE,
  Sept. 2013.

\bibitem{Julius2009}
A.~Julius and G.~Pappas.
\newblock Approximations of stochastic hybrid systems.
\newblock {\em IEEE Trans. on Automatic Control}, 54(6):1193--1203, June 2009.

\bibitem{Jr2010}
M.~{Mazo Jr}, A.~Davitian, and P.~Tabuada.
\newblock {PESSOA: towards the automatic synthesis of correct-by-design control
  software}.
\newblock In {\em Work-in-progress HSCC}, 2010.

\bibitem{sa2011}
S.~E.~Z. Soudjani and A.~Abate.
\newblock Adaptive gridding for abstraction and verification of stochastic
  hybrid systems.
\newblock In {\em Proc. of Quantitative Evaluation of Systems}, pages 59--68,
  Aachen, DE, 2011.

\bibitem{Tabuada2009b}
P.~Tabuada.
\newblock {\em {Verification and control of hybrid systems}}.
\newblock Springer US, Boston, MA, 2009.

\bibitem{tabuada2006linear}
P.~Tabuada and G.~J. Pappas.
\newblock {Linear time logic control of discrete-time linear systems}.
\newblock {\em Automatic Control, IEEE Transactions on}, 51(12):1862--1877,
  2006.

\bibitem{Vardi2006}
M.~Y. Vardi.
\newblock From philosophical to industrial logics.
\newblock In {\em Logic and Its Applications}, pages 89--115, Berlin,
  Heidelberg, 2009. Springer-Verlag.

\bibitem{Witsenhausen1971}
H.~Witsenhausen.
\newblock {Separation of estimation and control for discrete time systems}.
\newblock {\em Proceedings of the IEEE}, 59(11):1557--1566, 1971.

\bibitem{Zamani}
M.~Zamani, P.~Mohajerin~Esfahani, R.~Majumdar, A.~Abate, and J.~Lygeros.
\newblock Symbolic control of stochastic systems via approximately bisimilar
  finite abstractions.
\newblock {\em IEEE Trans. on Automatic Control,}, 59(12):3135--3150, Dec 2014.

\bibitem{zhang2005logic}
L.~Zhang, H.~Hermanns, and D.~N. Jansen.
\newblock Logic and model checking for hidden {M}arkov models.
\newblock In {\em Proc. on Formal Techniques for Networked and Distributed
  Systems}, pages 98--112. Springer, 2005.

\end{thebibliography}

\end{document}